\documentclass[a4paper,11pt]{article}
\usepackage[T1]{fontenc}
\usepackage[utf8]{inputenc}
\usepackage{lmodern}
\usepackage{amsmath}
\usepackage{epsfig}
\usepackage[margin=0.9in]{geometry}
\usepackage[english]{babel}
\usepackage{wrapfig}
\begin{document}
\title{New JLab/Hall A Deeply Virtual Compton Scattering results}

\author{{\slshape Maxime Defurne$^1$}\\ {\footnotesize for the Hall A DVCS collaboration}\\[1ex]
$^1$CEA, Centre de Saclay, IRFU/SPhN/LSN, F-91191 Gif-sur-Yvette, France\\}

\maketitle

\begin{abstract}
New data points for unpolarized Deeply Virtual Compton Scattering cross sections have been extracted from the E00-110 experiment at Q$^2$=1.9~GeV$^2$ effectively doubling the statistics available in the valence region. A careful study of systematic uncertainties has been performed.   
\end{abstract}

\begin{wrapfigure}{r}{0.4\textwidth} 
\begin{center}
\includegraphics[width=0.4\textwidth]{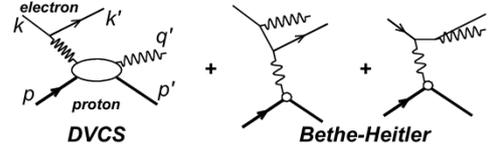}
\caption{\label{Fig:feyn}Lowest order QED diagrams for DVCS and Bethe Heilter processes. Defining $q=k-k'$, Q$^2$=$-|q|^2$ and $t$=$|p-p'|^2$. $x_{B}$ is given by $\frac{Q^2}{2p\cdot q}$.}
\end{center}
\end{wrapfigure}
Generalized Parton Distributions (GPDs) correlate the spatial and momentum distributions of partons inside the nucleon. They are nowadays the main way to study the orbital angular momentum of quarks via the Ji's sum rule. As GPDs are accessible through deep exclusive processes, a worldwide experimental program has been developped to study them~\cite{Guidal:2013rya}. Experiment E00-110 has been designed to investigate the electroproduction of photons ($ep\rightarrow ep\gamma$). Beam helicity dependent cross sections at $x_B$=0.36 and Q$^2$ = \{1.5,1.9,2.3\}~GeV$^2$ have been published by Munoz \emph{et al.} in 2006~\cite{MunozCamacho:2006hx}. An additional unpolarized cross section at the highest value of Q$^2$ was extracted at 2.3~GeV$^2$. Here we present the extraction of the unpolarized cross section at the intermediate Q$^2$= 1.9~GeV$^2$.     

\section{Phenomenological framework}
Photon electroproduction in the deep inelastic kinematics includes the coherent contribution of Bethe-Heitler, where the photon is emitted by the incoming or scattered electron, and Deeply Virtual Compton Scattering (DVCS) where the photon is emitted by the proton (see figure~\ref{Fig:feyn}). The amplitude for DVCS is parametrized by Compton form factors (CFF) which are complex integral of GPDs. The interference between these two processes makes the photon electroproduction a golden channel because it gives access to the real and imaginary parts of CFFs. Kumericki and Muller~\cite{Belitsky:2010jw} performed a Fourier expansion of the different contributions according to $\phi$, the angle between the leptonic and the hadronic plane. The information about the GPD is embedded in the Fourier coefficients of the DVCS amplitude and the interference term. The amplitude of photon electroproduction $A_{ep\rightarrow ep\gamma}$ is given by:
\begin{eqnarray}
|A_{ep\rightarrow ep\gamma}|^2&=&|A_{DVCS}|^2+|A_{BH}|^2+I_{BH/DVCS},\mbox{ with}\nonumber\\
|A_{DVCS}|^2&\propto&c_0^{DVCS}+\sum_{n=1}^{2} \Big(c_n^{DVCS} cos(n\phi) + s_n^{DVCS} sin(n\phi)\Big)\nonumber\\
I_{BH/DVCS}&\propto&c_0^{I}+\sum_{n=1}^{2} \Big(c_n^{I} cos(n\phi) + s_n^{I} sin(n\phi)\Big)\nonumber\\
\end{eqnarray}

Indeed $c_n^{DVCS}$ and $s_n^{DVCS}$ (respectively $c_n^{I}$ and $s_n^{I}$) are bilinear (respectively linear) combinations of CFFs. The amplitude of the Bethe Heitler is exactly known assuming a reliable parameterization of the form factors of the nucleon.
The beam helicity independent cross section is mostly sensitive to $\mathcal{HH}^*$ and Re$\mathcal{H}$, and the difference of beam helicity dependent cross sections to Im$\mathcal{H}$.

\section{Experimental setup}  
The experiment ran in the Hall A of Jefferson Laboratory~\cite{Alcorn:2004sb} in the spring of 2004, using the 80\%-polarized 5.75~GeV continuous electron beam provided by CEBAF impinging on a 15-cm long liquid hydrogen target. The left high resolution spectrometer was dedicated to the scattered electron detection.  

A dedicated electromagnetic calorimeter made of 11 $\times$ 12 = 132 lead fluoride blocks read by photomultiplier tubes (PMTs) was used to detect the outgoing photon. 

A recoil detector was built for the proton detection but it was demonstrated that a cut on the squared missing mass associated to the reaction $ep\rightarrow e\gamma X$ was enough to ensure the exclusivity. As the proton detector was limiting the acceptance, it was not used in this analysis.

\section{Subtraction of $\pi^0$ contamination}
In their center-of-mass frame, $\pi^0$ isotropically decay into two photons, emited back-to-back. While, in the laboratory frame, due to the directionality of the Lorentz boost, the decay photons share the energy asymetrically in most cases. As a result, one of them may get most of the energy and the other one almost nothing, impossible to detect because of the 1~GeV threshold imposed on the calorimeter. In that case, as exclusive $\pi^0$ have an energy close to the one of an exclusive photon, we will interpret it as an exclusive photon. 

To subtract this contamination, The sample of $\pi^0$'s whose two photons have been detected is used. Knowing their 4-momenta, we simulate their decay $N_{gen}$=5000 times thanks to a Monte Carlo simulation. Among the $N_{gen}$ decays, there are:
\begin{itemize}\setlength{\itemsep}{1mm}
\item $n_0$ events where none of the photons have been detected, or only one photon detected but with an associated missing mass not compatible with an exclusive photon event.
\item $n_2$ events where the two photons are detected.
\item $n_1$ events where one photon is detected with a missing mass compatible with an exclusive photon event.
\end{itemize}

For each of the $n_1$ decays, the kinematic variables $t$, $\phi$ are computed as if it was an exclusive photon event. Then this event is considered with the weight $\frac{1}{n_2}$ in the corresponding experimental bin. At the end of the day, the contamination is estimated in all the experimental bins.

 This method naturally includes the $\pi^0$ electroproduction cross section in the subtraction. Since it relies strongly on our ability to detect the two photons of the decay, we apply a geometrical cut on the calorimeter surface to remove its edges and corners. 
 
 \begin{figure}[!htp]
\begin{tabular}{cc}
\includegraphics[width=0.5\textwidth]{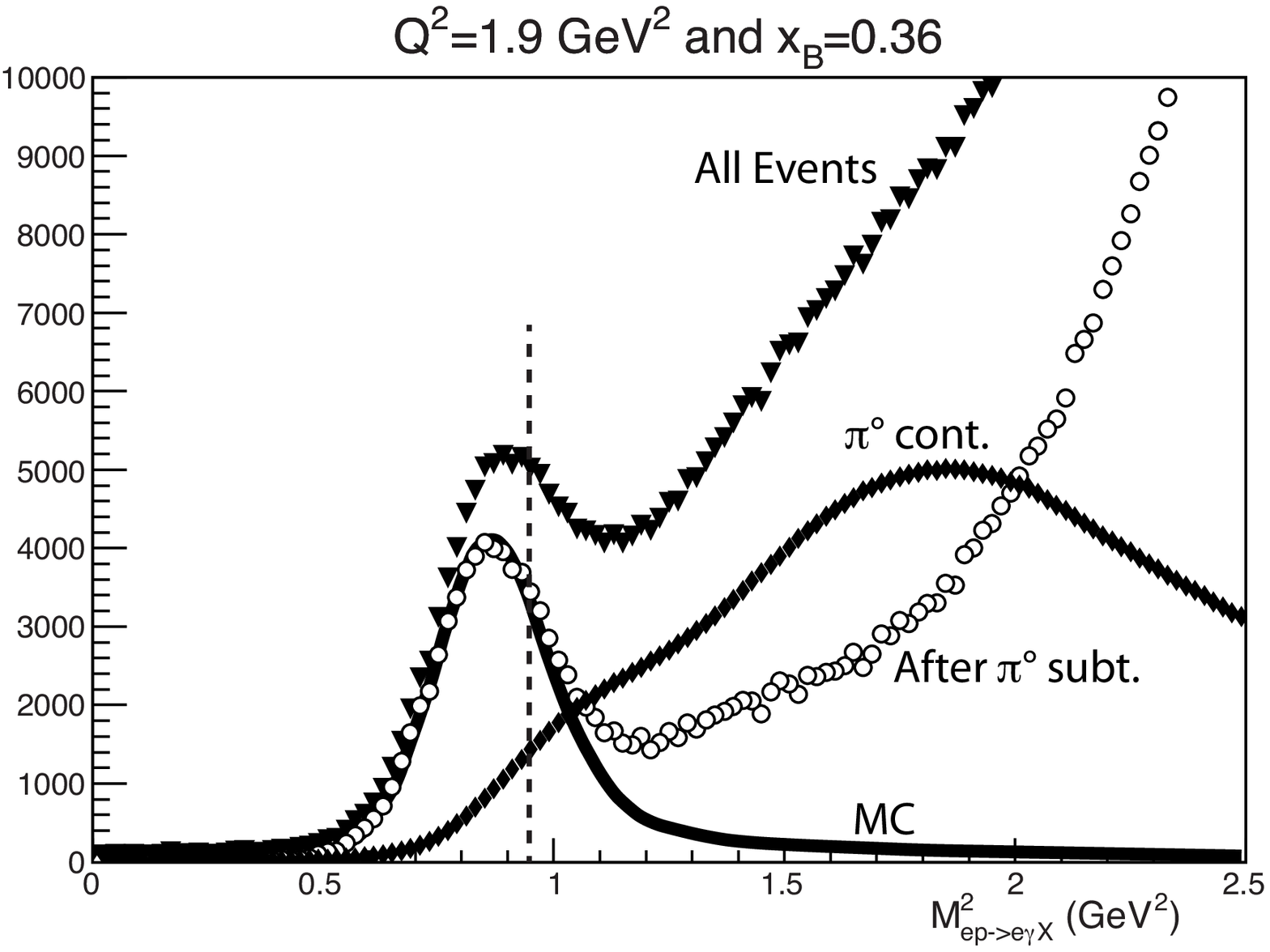} & \includegraphics[width=0.5\textwidth]{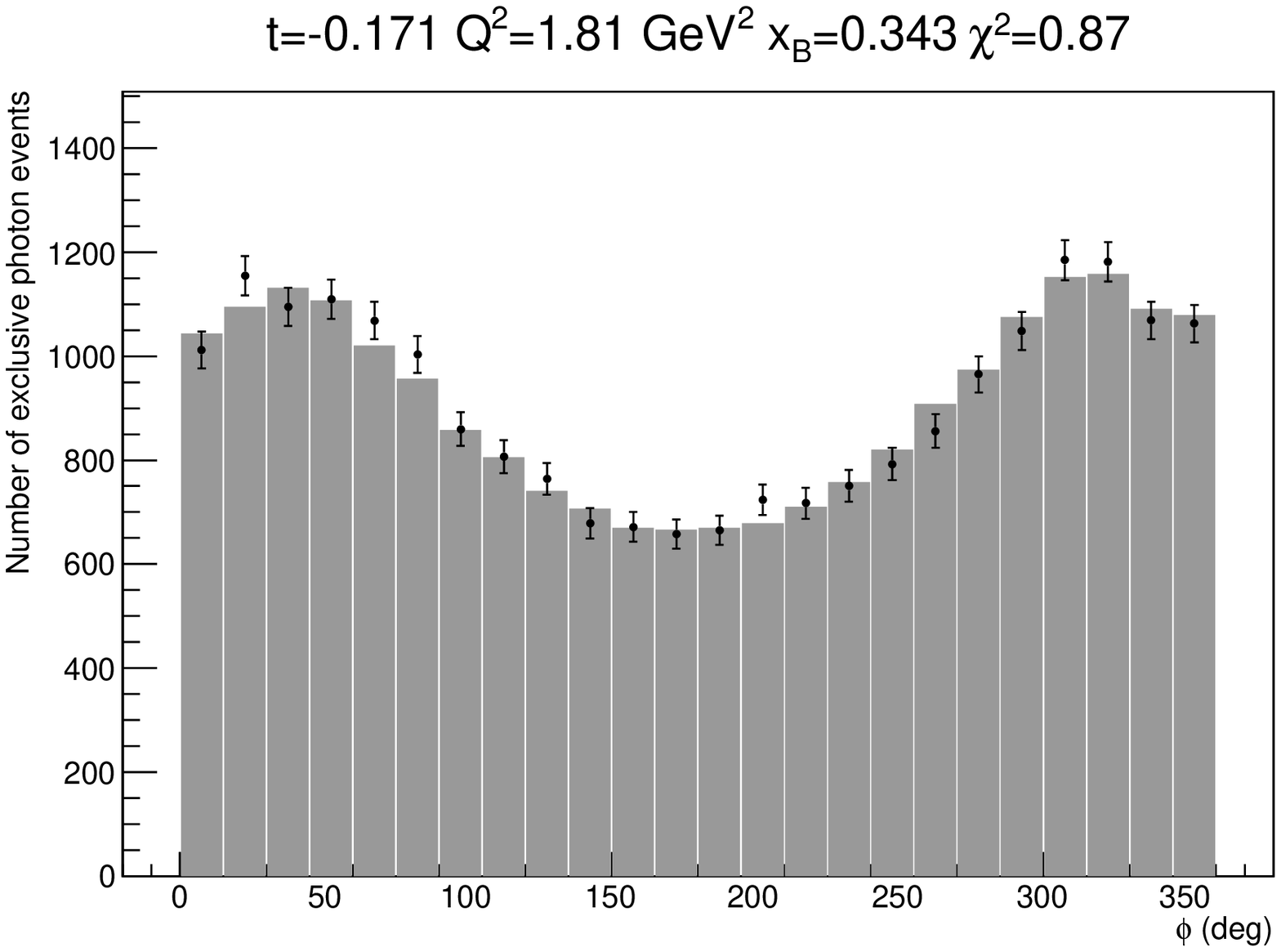} \\
\end{tabular}
\caption{Right: Missing mass spectrum associated to $ep\rightarrow e\gamma X$. To ensure exclusivity, we require a value below 0.95 GeV$^2$. The markers are the number of counts from experiment. The histograms represent the number of counts expected by the Monte Carlo simulation once the cross section has been extracted by the fitting procedure.}\label{Fig:MV}
\end{figure}

\section{Monte Carlo simulation} 
The Monte Carlo simulation has been upgraded to Geant4. Radiative corrections are applied following the method described in~\cite{Vanderhaeghen:2000ws}. Emission of soft photons from internal bremmstrahlung is handled using the equivalent radiator method.
 
Because of radiation damage, blocks close to the beam have a poorer energy resolution than the ones far from the beam. As a consequence, the exclusivity peak in the $M^2_{ep\rightarrow e\gamma X}$ will be larger close to the beam than far from it. Since binning in $t$ and $\phi$ translates into geometrical cuts in the calorimeter, it is vital to have a good match between the Monte Carlo and the experimental missing mass spectrum.

To estimate the error due to the exclusivity cut, we studied the cross section variations when changing the missing mass cut. 

\section{Cross section and CFF extraction}
 Using the formalism developped in ~\cite{Belitsky:2010jw}, we parameterize the cross section in terms of CFFs. However there are too many unknowns with respect to our data. By assuming twist-2 dominance and a sizeable $|DVCS|^2$ contribution (as hinted in~\cite{MunozCamacho:2006hx}), we end up using three parameters in order to fit each data bin in $\phi$ and $t$ (equation~\ref{eq:KM}). We studied 5 bins in $t$, each of them with 24 bins in $\phi$, giving a total number of bins $N_{bin}$=120.

To fit each of the 5 $t$-bins, we minimize the following $\chi^2$:
\begin{equation} \label{chi2}
\chi^2=\sum_{k=0}^{N_{bin}} \Bigg(\frac{N^{exp}_k-N^{sim}_k}{\sigma^{exp}_k}\Bigg)^2
\end{equation}
where $N_k^{exp}$ is the number of counts in bin $k$ from data after subtraction of contamination, and $\sigma^{exp}_k$ represents the statistical uncertainty on the number of counts in the bin $k$. $N^{sim}_k$ the number of counts in the bin $k$ expected with the Monte Carlo simulation and is given by: 
\begin{eqnarray}
N^{sim}_k&=&L \int_{\Phi_k}\frac{d^4\sigma}{d\Phi}d\Phi_k ,\\
\label{eq:KM} \frac{d^4\sigma}{d\Phi}&=& \frac{d^4\sigma_{BH}}{d\Phi}+\Gamma^{\mathcal{C}^{DVCS}_{unp}}\times \mathcal{C}^{DVCS}_{unp}+\Gamma^{\mathcal{C}^I(\mathcal{F})}\times Re\mathcal{C}^I(\mathcal{F})+\Gamma^{\mathcal{C}^I(\mathcal{F}_{eff})}\times Re\mathcal{C}^I(\mathcal{F}_{eff}) ,
\end{eqnarray}
with $L$ the integrated luminosity of the experiment and $\Phi_k$ the phase space of the experimental bin.

The coefficients $\Gamma$ in equation~\ref{eq:KM} are given by~\cite{Belitsky:2010jw} and depend on $\phi$, $t$, $x_B$ and Q$^2$. Their integral is performed using the Monte Carlo simulation and help us to take into account most of the kinematic dependences. Finally, by evaluating the coefficients $\Gamma$ at the vertex and applying selection cuts on the variables reconstructed by the detectors, we correct for bin migration. 

At the end of the day, we obtain unpolarized photon electroproduction cross sections at $x_B$=0.36 and Q$^2$=1.9 GeV$^2$. The photon electroproduction cross sections will be published in April 2015.

\begin{footnotesize}
\bibliographystyle{ieeetr}
\bibliography{arXiv}

\end{footnotesize}

\end{document}